\documentclass[reprint,superscriptaddress,citeautoscript,floatfix,longbibliography]{revtex4-1}
\usepackage[T1]{fontenc}
\usepackage[utf8]{inputenc}
\usepackage{times,siunitx,amsmath,amssymb,amsfonts,hyperref,graphicx}
\usepackage[capitalize]{cleveref}
\newcommand{\para}{{\mkern3mu\vphantom{\perp}\vrule depth 0pt\mkern3mu\vrule depth 0pt\mkern3mu}}
\begin{document}

\title{Dimensionality of the superconductivity in the transition metal pnictide WP}

\author{Angela Nigro}
\affiliation{Dipartimento di Fisica "E.R. Caianiello", Università degli Studi di Salerno, 84084 Fisciano (Salerno), Italy}
\affiliation{Consiglio Nazionale delle Ricerche CNR-SPIN, UOS Salerno, 84084 Fisciano (Salerno), Italy}

\author{Giuseppe Cuono}
\email{gcuono@magtop.ifpan.edu.pl}
\affiliation{International Research Centre Magtop, Institute of Physics, Polish Academy of Sciences, Aleja Lotnik\'ow 32/46, PL-02668 Warsaw, Poland}

\author{Pasquale Marra}
\email{pmarra@ms.u-tokyo.ac.jp}
\affiliation{Graduate School of Mathematical Sciences, The University of Tokyo, Komaba, Tokyo, 153-8914, Japan}
\affiliation{Department of Physics, and Research and Education Center for Natural Sciences, Keio University, Hiyoshi, Kanagawa, 223-8521, Japan}

\author{Antonio Leo}
\affiliation{Dipartimento di Fisica "E.R. Caianiello", Università degli Studi di Salerno, 84084 Fisciano (Salerno), Italy}
\affiliation{NANO\_MATES Research Centre for NANOMAterials and NanoTEchnology at Salerno University, Università degli Studi di Salerno, 84084 Fisciano (Salerno), Italy}
\affiliation{Consiglio Nazionale delle Ricerche CNR-SPIN, UOS Salerno, 84084 Fisciano (Salerno), Italy}

\author{Gaia Grimaldi}
\affiliation{Consiglio Nazionale delle Ricerche CNR-SPIN, UOS Salerno, 84084 Fisciano (Salerno),
Italy}
\affiliation{Dipartimento di Fisica "E.R. Caianiello", Università degli Studi di Salerno, 84084 Fisciano (Salerno), Italy}

\author{Ziyi Liu}
\affiliation{Beijing National Laboratory for Condensed Matter Physics and Institute of Physics, Chinese Academy of Sciences, Beijing 100190, China}

\author{Zhenyu Mi}
\affiliation{Beijing National Laboratory for Condensed Matter Physics and Institute of Physics, Chinese Academy of Sciences, Beijing 100190, China}

\author{Wei Wu}
\affiliation{Beijing National Laboratory for Condensed Matter Physics and Institute of Physics, Chinese Academy of Sciences, Beijing 100190, China}

\author{Guangtong Liu}
\affiliation{Beijing National Laboratory for Condensed Matter Physics and Institute of Physics, Chinese Academy of Sciences, Beijing 100190, China}
\affiliation{Songshan Lake Materials Laboratory, Dongguan, Guangdong 523808, China}

\author{Carmine Autieri}
\affiliation{International Research Centre Magtop, Institute of Physics, Polish Academy of Sciences, Aleja Lotnik\'ow 32/46, PL-02668 Warsaw, Poland}
\affiliation{Consiglio Nazionale delle Ricerche CNR-SPIN, UOS Salerno, 84084 Fisciano (Salerno), Italy}

\author{Jianlin Luo}
\affiliation{Beijing National Laboratory for Condensed Matter Physics and Institute of Physics, Chinese Academy of Sciences, Beijing 100190, China}
\affiliation{Songshan Lake Materials Laboratory, Dongguan, Guangdong 523808, China}
\affiliation{School of Physical Sciences, University of Chinese Academy of Sciences, Beijing 100190, China}

\author{Canio Noce}
\affiliation{Dipartimento di Fisica "E.R. Caianiello", Università degli Studi di Salerno, 84084 Fisciano (Salerno), Italy}
\affiliation{Consiglio Nazionale delle Ricerche CNR-SPIN, UOS Salerno, 84084 Fisciano (Salerno), Italy}

\begin{abstract}
We report theoretical and experimental results on the transition metal pnictide WP\@. 
The theoretical outcomes based on tight-binding calculations and density functional theory indicate that WP is a three-dimensional superconductor with an anisotropic electronic structure and nonsymmorphic symmetries.
On the other hand, magnetoresistance experimental data and the analysis of superconducting fluctuations of the conductivity in external magnetic field indicate a weakly anisotropic three-dimensional superconducting phase.
\end{abstract}

\maketitle

\section{Introduction}

The discovery of superconductivity under external pressure in the chromium arsenide CrAs stimulated considerable efforts in the quest of superconductivity in other binary pnictides at ambient pressure~\cite{Wu14,Kotegawa14,Wu10}. 
CrAs belongs to the family of transition metal pnictides with chemical formula MX (with M~=~transition metal and X~=~P, As, Sb), and it has an orthorhombic MnP-type crystal structure at ambient conditions. 
Soon after this discovery, a new member of the same family, the transition metal phosphide MnP, has been grown~\cite{Cheng15}. 
Both MnP and CrAs become superconducting under external pressure and exhibit a similar temperature--pressure phase diagram with a superconducting dome~\cite{Wu14,Kotegawa14,Cheng15} and the presence of magnetic phases~\cite{Cuono21APPA} which can coexist with superconductivity~\cite{Wu14,Kotegawa14,Cheng15,Chen19,Noce20}.

\begin{figure}
\centering
\includegraphics[width=\columnwidth]{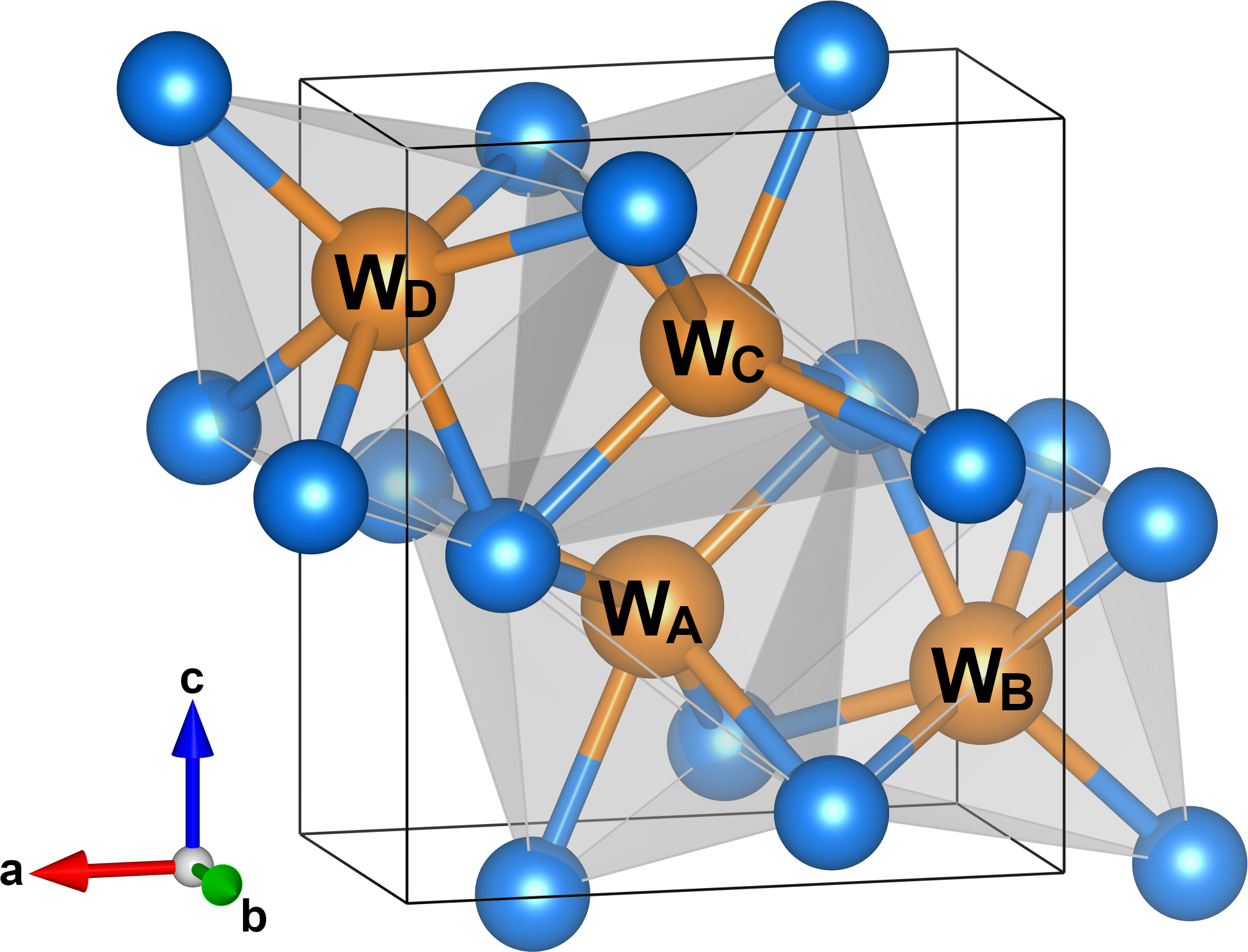}
\setlength{\belowcaptionskip}{-12pt}\setlength{\abovecaptionskip}{-1pt}
\caption{
The orthorhombic crystal structure of tungsten phosphide WP with space group Pnma. 
Orange and blue spheres indicate W and P ions, respectively, with nonequivalent lattice positions of the W ions labeled as $\mathrm{W}_A$, $\mathrm{W}_B$, $\mathrm{W}_C$, $\mathrm{W}_D$. 
Face-sharing WP$_6$ octahedra are shaded in gray. 
}
 \label{Structure}
\end{figure}

More recently, a new superconductor of the same series has been produced, namely the tungsten phosphide WP, with bulk superconductivity appearing at \SI{0.84}{K} at ambient pressure~\cite{Liu19}. 
So far, WP is the only known example of a 5$d$ transition metal phosphide with
no long-range magnetic order.
The primitive cell of WP contains four W and four P atoms, with each W atom surrounded by six nearest-neighbor P atoms, and located at the center of the face-sharing WP$_6$ octahedra~\cite{Liu19}, shown in \cref{Structure}. 
Four of the six bonds are inequivalent due to the space group anisotropy~\cite{Liu19}. 
In this compound, the spatial extension of the W-5$d$ orbitals induces a large overlap and a strong coupling with the neighboring $p$-orbitals, 
resulting in a distortion of the crystal structure more pronounced compared to that of CrAs and MnP~\cite{Liu19,Cuono19}. 
Moreover, the spin-orbit coupling of W-5$d$ electrons is stronger than that of 3$d$ electrons of CrAs and MnP~\cite{Cuono19}. 
In particular, the 
3$d$ materials display strong electron correlations, narrow bandwidths, and robust magnetism.
On the other hand, the 5$d$ materials exhibits increased hybridization, more diffuse orbitals, and a strong spin-orbit coupling competing with magnetic, crystal-field, many-body Coulomb, and other interactions
leading to novel and exotic behaviors~\cite{Kim09,Cao18}. 
Moreover, the relativistic shifts in orbital energies, combined with spin-orbit and bandwidth effects, drive band inversions leading to topological phases and enhanced Rashba splittings.

This class of materials exhibits a nonmagnetic phase with no long-range magnetic order with a Fermi surface consisting of hole-like branches with two-dimensional (2D) dispersion, together with electron-like branches with a full three-dimensional (3D) character~\cite{Autieri17,Cuono19}. 
Therefore, it is legitimate to ask what is the dimensionality of transport in these systems in the superconducting phase. It is well-known that this depends on the structure of the order parameter in the $k$-space associated with the points of the Fermi surface. 
In particular, when the hole-like surfaces exhibit a vanishing order parameter, the system can be considered a 3D isotropic superconductor with the upper critical field $H_{c2}$ approximately equal in all directions.
On the other hand, if the electron-like surface exhibits a vanishing order parameter, then the system can be considered an anisotropic superconductor with a strong 2D character, with the in-plane upper critical field $H_{c2}^\para$ much larger than the one in the perpendicular direction $H_{c2}^\perp$.
Finally, when the superconductivity comes from both branches of the Fermi surface, the system will be an anisotropic 3D superconductor with $H_{c2}^{\para}/H_{c2}^{\perp}\neq 1$.

Although many interesting studies have been so far reported on these and other similar compounds, both theoretical~\cite{Autieri17,Autieri17b,Autieri18,Cuono19a,Niu17,Daido19,Wysokinski19,Gercsi10,Continenza01,Edelmann17,Cuono18,Cuono19b,Cuono20,Cuono20b} and experimental~\cite{Shen16,Wang16,Matsuda18,Sen19,Guo18,Nigro19,Keller15,Khasanov15,Kotegawa15,Park19,Wu20}, only a limited amount of information on WP single crystal system is until now available~\cite{Liu19}. 
The aim of this paper is to give a contribution to the investigation of the properties of WP compound,
providing electronic structure calculations and a consistent description of resistivity measurements from experiments carried out on WP single crystal samples. 
In particular, we will focus on isothermal magnetoresistance $R(H)$ measurements performed at different magnetic field directions with respect to the $a$-axis and on the resistivity measurements $R(T)$ at different applied magnetic fields, which reveal the superconducting fluctuations around the critical temperature $T_c$.
 
We notice that magnetoresistance measurements performed as a function of the applied field direction will give remarkable information about the properties of WP\@. 
Indeed, the comparison between theoretical models for superconductors with 2D and 3D character~\cite{Tinkham63,Ketterson99} will provide an indication about the anisotropy of the upper critical field and the effective mass, revealing the dimensional behavior of WP\@.

Furthermore, the analysis of the temperature dependence of the resistivity fluctuations under the application of a magnetic field, through the scaling procedure obtained theoretically by Ullah and Dorsey model~\cite{Ullah90}, will help to discriminate again between 2D and 3D characters in WP\@.

Here, we will prove that these experimental outcomes, together with the theoretical background, give important insights on the properties of WP, unveiling relevant microscopic aspects of this material, suggesting that this compound can be considered a 3D anisotropic superconductor. 
The paper is organized as follows: 
In the next section, we will describe the experimental methods adopted to synthesize WP;
In section III, we present the theoretical calculations and the experimental data about magnetoresistance and paraconductivity measurements;
Finally, the last section will be devoted to the discussion of the results and the conclusions.

\section{Sample preparation method and experimental details}

We grew high-quality WP needle-like single crystals by chemical vapor transport method, which has also been used to grow other pnictides~\cite{Motojima85,Yang15,Liu16,Liu19}. 
The starting WP polycrystalline powders and iodine were placed in a quartz tube sealed under high vacuum in a two-zone furnace with a temperature gradient from \SI{1100}{\celsius} to \SI{1050}{\celsius} for one week. 
Then the sample temperature was raised up to \SI{1150}{\celsius} in a one-month period. 
Using this method, we grew single WP single crystals with a typical dimension of $0.1\times0.1\times\SI{4.0}{mm^3}$.

\begin{figure}
\centering
\includegraphics[width=\columnwidth]{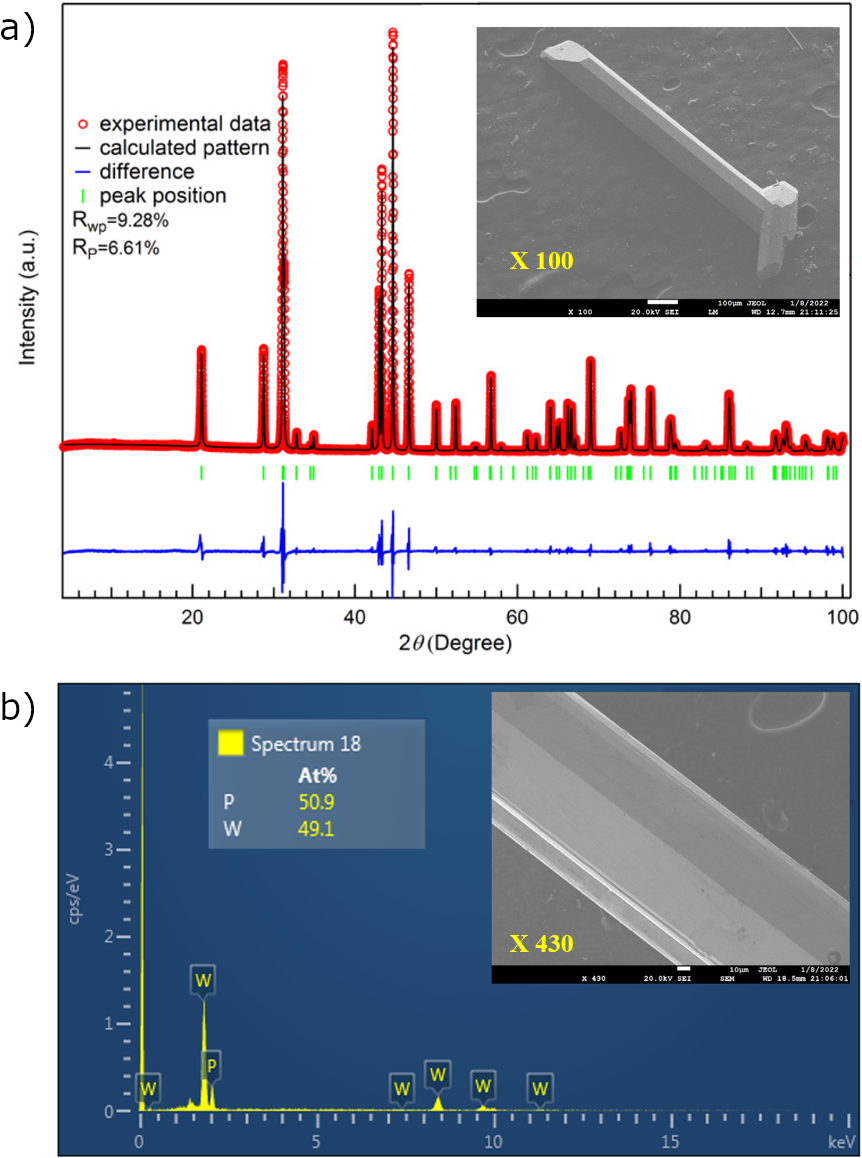}
\setlength{\belowcaptionskip}{-12pt}\setlength{\abovecaptionskip}{-1pt}
\caption{
(a) 
Room-temperature X-ray diffraction patterns and the Rietveld refinement of WP. 
Open circles, solid line, and lower solid line represent experimental, calculated, and difference XRD patterns, respectively. 
The inset shows the SEM image for WP single crystal. 
(b) 
EDX result of WP single crystal. 
The average ratio of the elements is close to the $1:1$ stoichiometry of the compound. 
The inset shows the SEM image for WP single crystal ($\times$430 magnification).
}
\label{XRDEDX}
\end{figure}

\Cref{XRDEDX}(a) shows the 
powder x-ray diffraction (XRD) data on WP at room temperature. 
The data show that WP crystallizes in a MnP-type orthorhombic structure (space group Pnma, No.~62) with lattice parameters $a=\SI{0.57222(6)}{nm}$, $b=\SI{0.32434(9)}{nm}$, and $c=\SI{0.62110(6)}{nm}$. 
The $b$-axis direction is parallel to the longest direction of the sample. 
The energy-dispersive x-ray spectroscopy (EDX) was performed to check the chemical composition of the grown single crystals.
\Cref{XRDEDX}(b) shows 
the typical EDX spectrum of an individual crystal. 
Only two elements, W and P, are detected. 
The average ratio of the elements at different locations in the crystals is $50.9:49.1$, which is close to the $1:1$ stoichiometry of the compound.
Further details on the fabrication procedure and the structural, compositional, and transport characterizations are reported elsewhere~\cite{Liu19}. 
The electrical resistance measurements below \SI{2}{K} were performed by the standard four-probe technique in a top-loading Helium-3 refrigerator with a superconducting magnet with fields up to \SI{15}{T}.

\section{Theoretical and experimental results}

In this section, we will report and discuss the theoretical simulation by employing density functional theory (DFT) and tight-binding approach, and the experimental results on the superconducting fluctuations of the conductivity of WP single crystals. 
We notice that the structural, elastic, mechanical, electronic, vibrational, and superconducting properties of WP have been discussed by using the outcomes from first-principle calculations based on DFT, also considering the spin-orbit coupling, in Ref.~\onlinecite{Tayran19}. 

\begin{table}[t]
\centering
\setlength{\tabcolsep}{10pt}
\renewcommand{\arraystretch}{1.5} 
\begin{tabular}{l|r}
{Parameters} \qquad & {Values}\\
 \hline
 $t^{100}_{\mathrm{W}_A\mathrm{W}_A}$ & $-0.284$\\
 $t^{200}_{\mathrm{W}_A\mathrm{W}_A}$ & $0.009$\\
 $t^{300}_{\mathrm{W}_A\mathrm{W}_A}$ & $-0.004$\\
 $t^{010}_{\mathrm{W}_A\mathrm{W}_A}$ & $-0.195$\\
 $t^{020}_{\mathrm{W}_A\mathrm{W}_A}$ & $0.081$\\
 $t^{030}_{\mathrm{W}_A\mathrm{W}_A}$ & $-0.053$\\
 $t^{001}_{\mathrm{W}_A\mathrm{W}_A}$ & $0.049$\\
 $t^{002}_{\mathrm{W}_A\mathrm{W}_A}$ & $-0.032$\\
 $t^{003}_{\mathrm{W}_A\mathrm{W}_A}$ & $0.004$\\
 $t^{100}_{\mathrm{W}_A\mathrm{W}_B}$ & $0.082$\\
 $t^{001}_{\mathrm{W}_A\mathrm{W}_C}$ & $0.001$\\
 $t^{00\bar{1}}_{\mathrm{W}_A\mathrm{W}_C} $ & $0.319 $\\
\end{tabular}
\caption{
Values of the hopping parameters of the tight-binding minimal model (energy units in eV).
}
\label{tab2}
\end{table}

\subsection{Theoretical calculations}

The WP is a system which exhibits nonsymmorphic symmetries. 
It is well-known that the nonsymmorphic symmetries in the Pnma structure are responsible for exotic topological behaviors like the topological nonsymmorphic crystalline superconductivity~\cite{Daido19}, 2D Fermi surface topology~\cite{Cuono19}, Dirac topological surface states~\cite{Gao20}, and topologically-driven linear magnetoresistance~\cite{Campbell20}.
A detailed analysis of the effects of the nonsymmorphic symmetries on the fermiology of this compound and a tight-binding minimal model fitted to the DFT band structure has been reported elsewhere~\cite{Cuono19}.
Here, we provide further investigation using a low-energy tight-binding model in order to calculate the hopping parameters at the Fermi level and give an indication about the dimensionality of the energy spectra.
We restrict to the representative subspace of one $d$-orbital for every W atom of the unit cell and consider only non-vanishing projected W-W hopping amplitudes. 
In \cref{tab2} we report the values of the parameters of the tight-binding minimal model, in which we have included the nearest-neighbor hopping terms along $x$, $y$ and $z$ directions. 
The parameters $t_{\alpha_i,\alpha_j}^{lmn}$ corresponds to the hopping amplitudes between sites $\alpha_i$ and $\alpha_j$ (where $i,j=\mathrm{W}_A,\mathrm{W}_B,\mathrm{W}_C,\mathrm{W}_D$ as in \cref{Structure}) along the direction $l\mathbf{x} + m\mathbf{y} + n\mathbf{z}$~\cite{Cuono19}.

\begin{figure}
\centering
\includegraphics[width=\columnwidth]{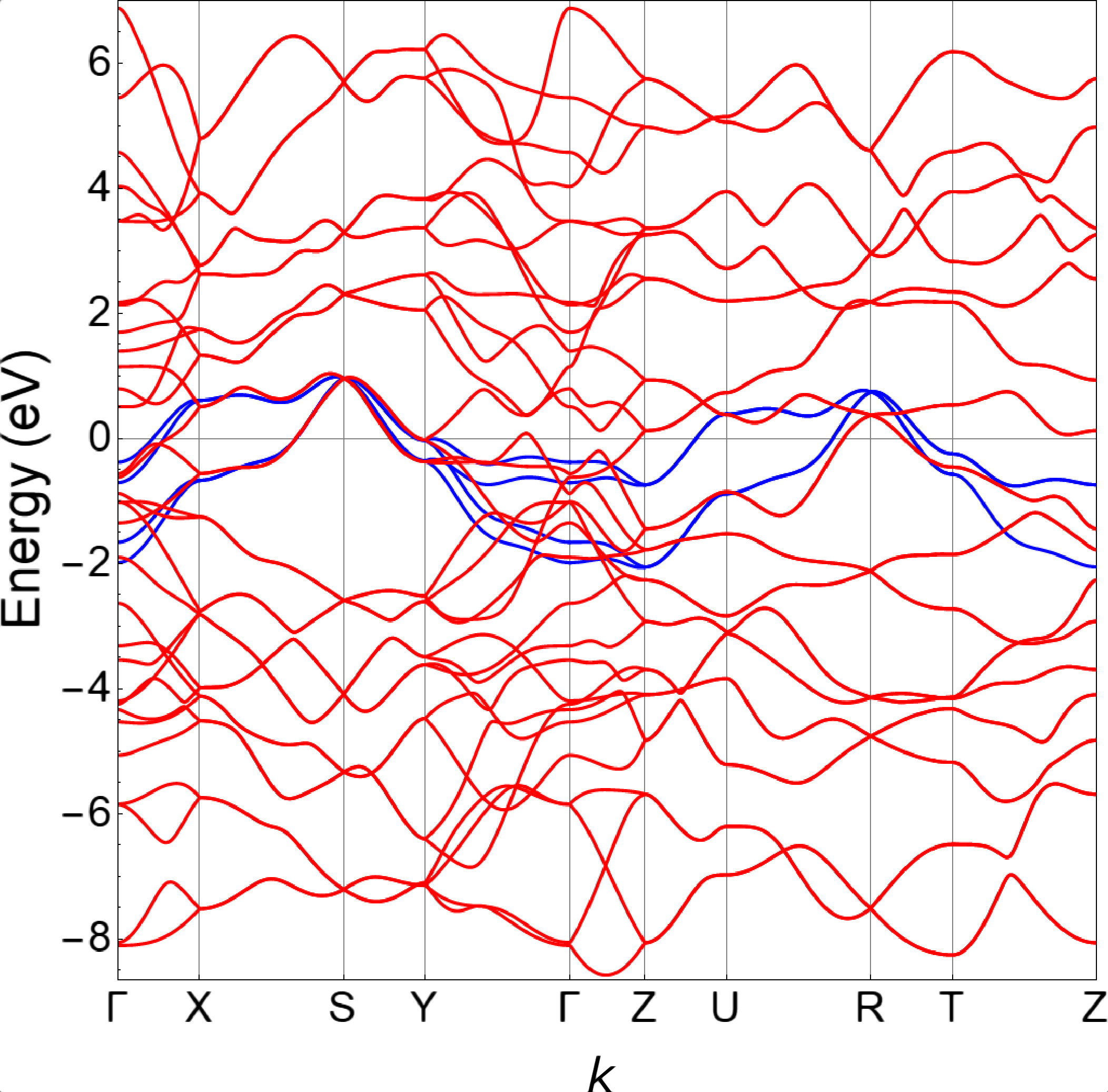}
\setlength{\belowcaptionskip}{-12pt}\setlength{\abovecaptionskip}{-1pt}
\caption{
Fit of the DFT bands (red lines) using the tight-binding model (blue lines) along the high-symmetry path of the orthorhombic Brillouin zone. 
The Fermi level is at zero energy.
}
\label{minimal1}
\end{figure}

\begin{figure}
\centering
\includegraphics[width=\columnwidth]{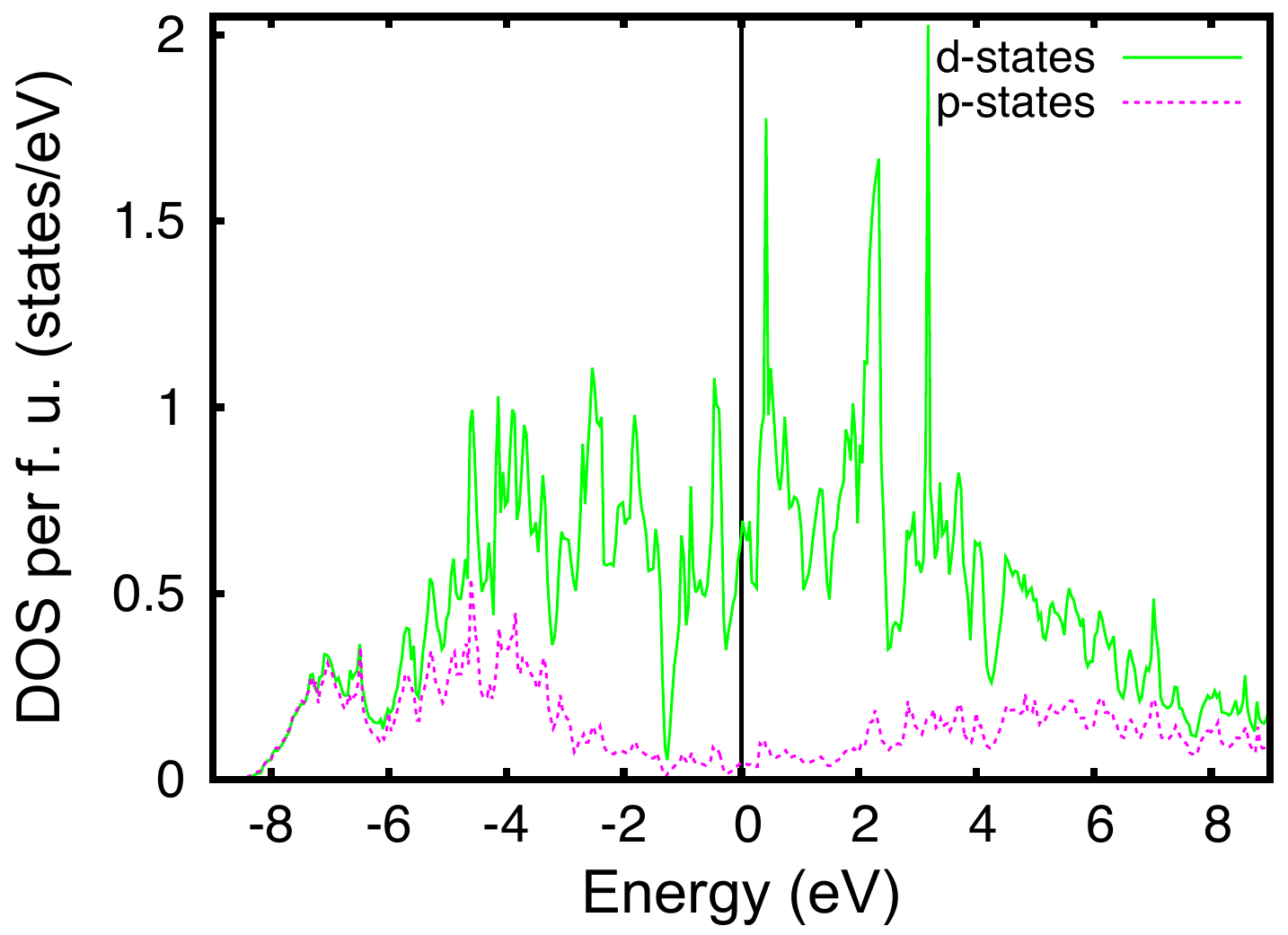}
\setlength{\belowcaptionskip}{-12pt}\setlength{\abovecaptionskip}{-1pt}
\caption{ 
Partial density of states relative to the W-5$d$ states (continuous line) and P-3$p$ states (dotted line). 
The $d$ states are predominant close to the Fermi level, while the $p$ states are far from the Fermi level, which is set at zero energy.
}
\label{DOS}
\end{figure}

From the examination of \cref{tab2}, we note that the dominant parameters along the $x$, $y$, and $z$ directions are $t^{100}_{\mathrm{W}_A\mathrm{W}_A}=t_{x}= \SI{-0.284}{eV}$, $t^{010}_{\mathrm{W}_A\mathrm{W}_A}=t_{y}= \SI{-0.195}{eV}$, and $t^{00\bar{1}}_{\mathrm{W}_A\mathrm{W}_C}=t_{z}= \SI{0.319}{eV}$. 
Moreover, the values of these parameters have the same order of magnitude, which indicates that the WP is fully 3D with moderate anisotropy.
For completeness, in \cref{minimal1} we show the fit of the DFT bands using the tight-binding minimal model along the high-symmetry path of the orthorhombic Brillouin zone. From this fit we infer that the model well captures all the symmetries along the high-symmetry lines of the Brillouin zone.
The orbital characters of the bands can be revealed by the partial density of states, shown in \cref{DOS}. 
P-3$p$ states dominate in the range [$-$8.5, $-$6] eV, while W-5$d$ states dominate in the range [$-$6, 4] eV.
Finally, above \SI{4}{eV} there is a mixing between P-3$p$, P-4$s$, and W-6$s$ states. 
Nevertheless, we notice that one cannot entirely decouple the W-5$d$ from the P-3$p$ states close to the Fermi level due to their strong hybridization~\cite{Cuono19}.

\begin{figure}
\centering
\includegraphics[width=\columnwidth]{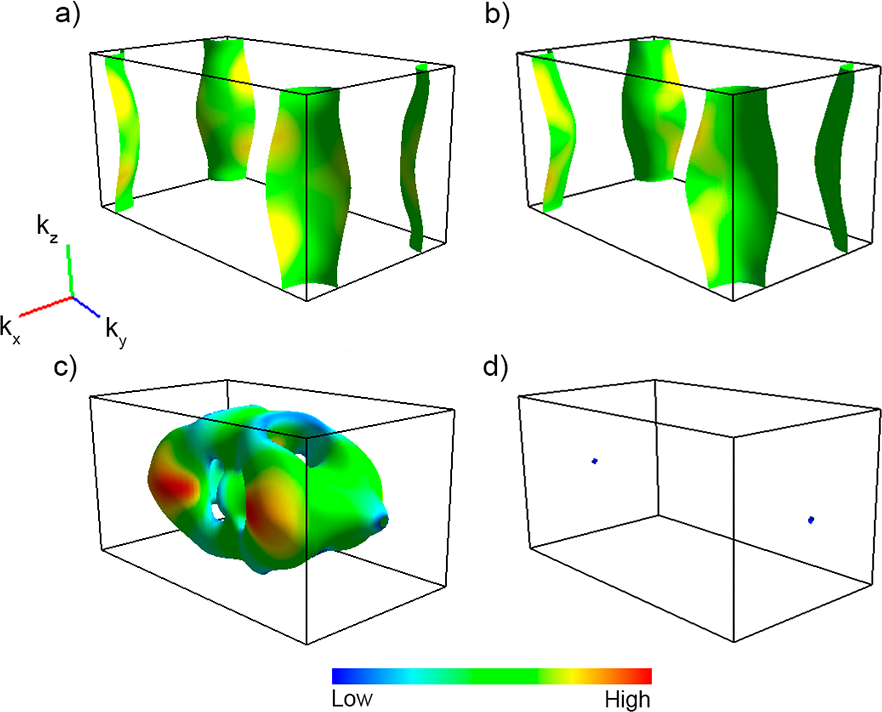}
\setlength{\belowcaptionskip}{-12pt}\setlength{\abovecaptionskip}{-1pt}
\caption{
Fermi surface of WP in the first Brillouin zone with spin-orbit coupling. 
In panels (a)-(d) we show the contributions of the four different bands that cut the Fermi level. 
The color code denotes the Fermi velocity.
}
\label{FS}
\end{figure}

Other evidence of the 3D behavior of the WP is provided by the analysis of the Fermi surface obtained through DFT calculations.
The DFT calculations have been performed using the VASP package~\cite{Kresse96}, treating the core and the valence electrons within the Projector Augmented Wave method~\cite{Kresse99} with a \SI{400}{eV} cutoff for the plane-wave basis.
The obtained Fermi surface is formed by a 3D branch around the center of the Brillouin zone and two hole-like 2D sheets centered around the SR high-symmetry line at ($k_x, k_y$)= ($\pi$,$\pi$) in the $ab$ plane 
(see the path for the orthorhombic Brillouin zone, space group 62, reported in Ref.~\onlinecite{Setyawan10}).
The presence of two hole-like 2D sheets is favored by nonsymmorphic symmetries~\cite{Cuono19}. 
Using the Fermisurfer code~\cite{Kawamura19}, we show in \cref{FS} the Fermi surface of the WP in the normal phase. 
The various colors indicate the different Fermi velocities, as shown in the color bar legend, with the highest Fermi velocities coming from the central 3D surface.

We see that the Fermi surface is formed by the four bands that cut the Fermi level, as shown in panels (a)-(d), 
suggesting that WP is an anisotropic 3D metal.
However, we notice that in the superconducting phase the degree of anisotropy could change because the anisotropy does not depend only on the bare electron band structure but also on the superconducting coupling~\cite{Tayran19}. 
Interestingly, a similar configuration in a borocarbide compound gives rise to a larger superconducting coupling within the 3D branches with respect to the 2D sheets of the Fermi surface~\cite{Kawamura17}.
Hence, we can speculate that WP is an anisotropic material with a normal state anisotropy larger than the superconducting phase anisotropy, due to a larger contribution to the superconducting order parameter from the 3D electron-like branch compared to the hole-like 2D sheets~\cite{Kawamura17}. 

\subsection{Experimental results}

A possible way to determine the dimensionality of a superconductor is represented by the study of the angular dependence of the upper critical field $H_{c2}$. 
It is well known indeed that this measurement may be indicative of a superconducting state with 2D or anisotropic 3D character. 
Within the Tinkham model, the behavior of the $H_{c2}(\theta)$, for a superconductor with 2D character, exhibits the following angular dependence~\cite{Tinkham63,Grimaldi18}
\begin{equation}
\left |\frac{H_{c2}(\theta) \sin(\theta)} {H_{c2}^{\para}}\right |+\left (\frac{H_{c2}(\theta) \cos(\theta)} {H_{c2}^{\perp}}\right )^2=1\, .
\end{equation}
On the other hand, for a superconductor with 3D character, and within the Ginzburg-Landau theory, the angular dependence of the upper critical field is written as~\cite{Ketterson99}
\begin{equation}
\left (\frac{H_{c2}(\theta) \sin(\theta)} {H_{c2}^{\para}}\right )^2+\left (\frac{H_{c2}(\theta) \cos(\theta)} {H_{c2}^{\perp}}\right )^2=1\, ,
\end{equation}
where $H_{c2}^\para$, $H_{c2}^\perp$, being the critical fields measured respectively in the parallel and perpendicular direction with respect to the sample surface, and 
$H_{c2}(\theta)$ the critical field measured at an angle $\theta$ with respect to the normal to the sample surface.

\begin{figure}
\centering
\includegraphics[width=\columnwidth]{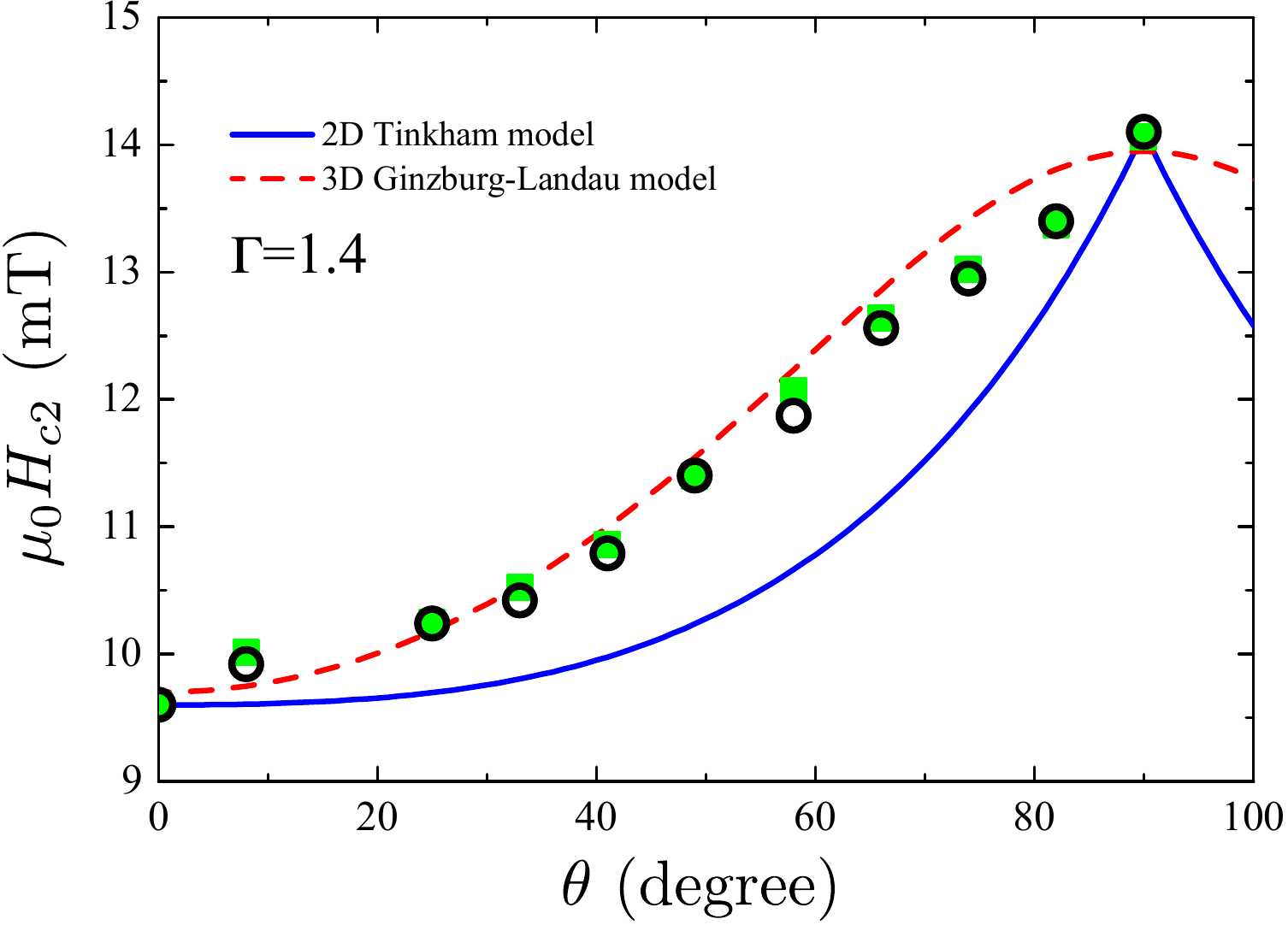}
\setlength{\belowcaptionskip}{-12pt}\setlength{\abovecaptionskip}{-1pt}
\caption{
Angular dependence of the superconducting upper critical field $H_{c2}$ at \SI{0.3}{K}. 
Black circles and green squares represent the measured transition fields (defined by the 90\% criterion) with positive and negative field polarity, respectively. 
The lines are the theoretical fits to the experimental data for the angular dependence of the critical field.
Red lines represent the theoretical dependencies according to the Ginzburg-Landau model for a superconductor with 3D character and anisotropic effective mass, blue lines to the Tinkham model for a superconductor with 2D character.
}
\label{GL3D}
\end{figure}

In our experiments, we find the anisotropy in the upper critical field $H_{c2}(\theta)$ when the field angle is rotated away from the $a$-axis, 
which corresponds to the axis perpendicular to the sample surface.
Extracting the values of $H_{c2}(\theta)$ from the resistive transition at several angles, with $T_c$ chosen at the 90\% of normal state resistance, we infer the full $H_{c2}(\theta)$ plot as given in \cref{GL3D}. 
As it can be observed, the $H_{c2}(\theta)$ experimental data are much better described by the anisotropic Ginzburg-Landau theory (red line), suggesting an anisotropic 3D environment for the superconductivity in WP\@. 
Interestingly, from these data we are able to infer also the degree of the anisotropy $\Gamma$ looking at the following ratio
\begin{equation}
\Gamma=\frac{H_{c2}^{\para}}{H_{c2}^{\perp}}\, ,
\end{equation}
We find that $\Gamma=1.4$ from which we may also estimate the effective mass ratio as ${m_{\perp}^*}/{m_{\para} ^*}=\Gamma^2$; 
This ratio is ${m_{\perp}^*}/{m_{\para} ^*} \simeq 2$, suggesting a moderate mass anisotropy. 

Since $H_{c2}^{\para}/H_{c2}^{\perp}>1$, we infer that WP may be considered a 3D anisotropic superconductor, and thus we expect that both electron and hole-like branches of the Fermi surfaces contribute to superconductivity.
For completeness, we notice that 3D anisotropic superconductivity has been intensively investigated in the last years because of its deviations from BCS theory even in superconductors with electron-phonon coupling~\cite{Kawamura17,Aperis15}.

The study of thermal fluctuation effects turns out to be another experimental tool to identify 3D rather than 2D thermal fluctuations. Moreover, it may offer several hints to understand relevant properties of WP, such as the occurrence of pronounced dissipation in the mixed state, detrimental for applications, and to provide essential information about the nature of the superconducting state.

It is well known that the understanding of superconducting fluctuations of conductivity around the transition temperature, in the presence of an applied field, requires a rather complex analysis. 
However, in a sufficiently high magnetic field, the paired quasiparticles are confined within the lowest Landau level (LLL) and, consequently, transport is restricted to the field direction.
In this case, the effective dimensionality of the system is reduced and the effect of fluctuations becomes more important. Specifically, the width of the temperature range around $T_c$, for a measurable excess conductivity, increases with the applied magnetic field as the in-field Ginzburg number $G_i(H)$ given by
\begin{align}
G^\mathrm{3D}_i(H)&=G^\mathrm{3D}_i(0)^{1/3}\left [ \frac{2H}{H_{c2}(0)} \right ]^{2/3} \, ,\\ 
G^\mathrm{2D}_i(H)&=G^\mathrm{2D}_i(0)^{1/2}\left [ \frac{2H}{H_{c2}(0)} \right ]^{1/2}\, ,
\label{GH}
\end{align}
for a superconductor with 3D and 2D character, respectively. 
Here, $H_{c2}(0)=- T_c dH_{c2} /dT |_{T=T_c}$ is the zero-temperature Ginzburg-Landau upper critical field, whereas $G^\mathrm{3D,2D}_i(0)$ are the zero-field Ginzburg numbers given by
\begin{equation}
G^\mathrm{3D}_i(0)=\frac{1}{2} \left ( \frac{k_B T_c}{E_c} \right )^2,\qquad
G^\mathrm{2D}_i(0)=\frac{k_B T_c}{E_F}\, .
\label{G0}
\end{equation}
In these formulas, $E_F$ is the Fermi energy, $E_c$ is the condensation energy within a coherence volume given by $E_c=(B^2_c(0)/2\mu_0)(\xi^3_\para(0)/\Gamma(0))$, $B_c(0)$ is the zero temperature thermodynamic critical field, $\xi_\para(0)$ is the zero temperature in-plane coherence length, and $\mu_0$ is the vacuum magnetic permeability~\cite{Blatter94,Larkin05,Ikeda89,Ikeda90,Eley17,Lee17,Tesanovic94}.

Ullah and Dorsey calculated the fluctuation conductivity using the LLL approximation and the self-consistent Hartree approximation, including contributions up to the quartic term in the free energy~\cite{Ullah90}. 
The resulting scaling law for the conductivity in the magnetic field, in terms of unspecified scaling functions $f_\mathrm{3D}$ and $f_\mathrm{2D}$ is valid for the 3D and 2D cases, respectively. 
The expression for the fluctuations of the conductivity $\sigma$ are
\begin{align}
\Delta \sigma_\mathrm{3D} (H)&=\left [ \frac{T^2}{H} \right ]^{1/3}
f_\mathrm{3D} \left [B\, \frac{T-T_c(H)}{(TH)^{2/3}} \right ]\, ,
\label{UD3D}
\\
\Delta \sigma_\mathrm{2D} (H)&=\left [ \frac{T}{H} \right ]^{1/2}
f_\mathrm{2D} \left [ A \, \frac{T-T_c(H)}{(TH)^{1/2}} \right ]\, ,
\label{UD2D}
\end{align}
for the 3D and 2D cases, respectively. 
In these expressions, known as the Ullah-Dorsey scaling law equations, $A$ and $B$ are appropriate constants characterizing the material.

These scaling laws 
describe the behavior of a large class of materials, including
amorphous low-$T_c$ superconductors~\cite{Theunissen97,Urbach94,Nikulov95},
high-$T_c$ cuprate superconductors, 
where these effects are much larger above a characteristic field $\mu_0H_{LLL}$ of the order of few teslas~\cite{Welp91,Costa97,Costa01}, and iron-based superconductors, 
with a measured field $\mu_0H_{LLL}=\SIrange{6}{8}{T}$~\cite{Pallecchi09,Pandya10,Pandya11,Marra12}.

\begin{figure}
\centering
\includegraphics[width=\columnwidth]{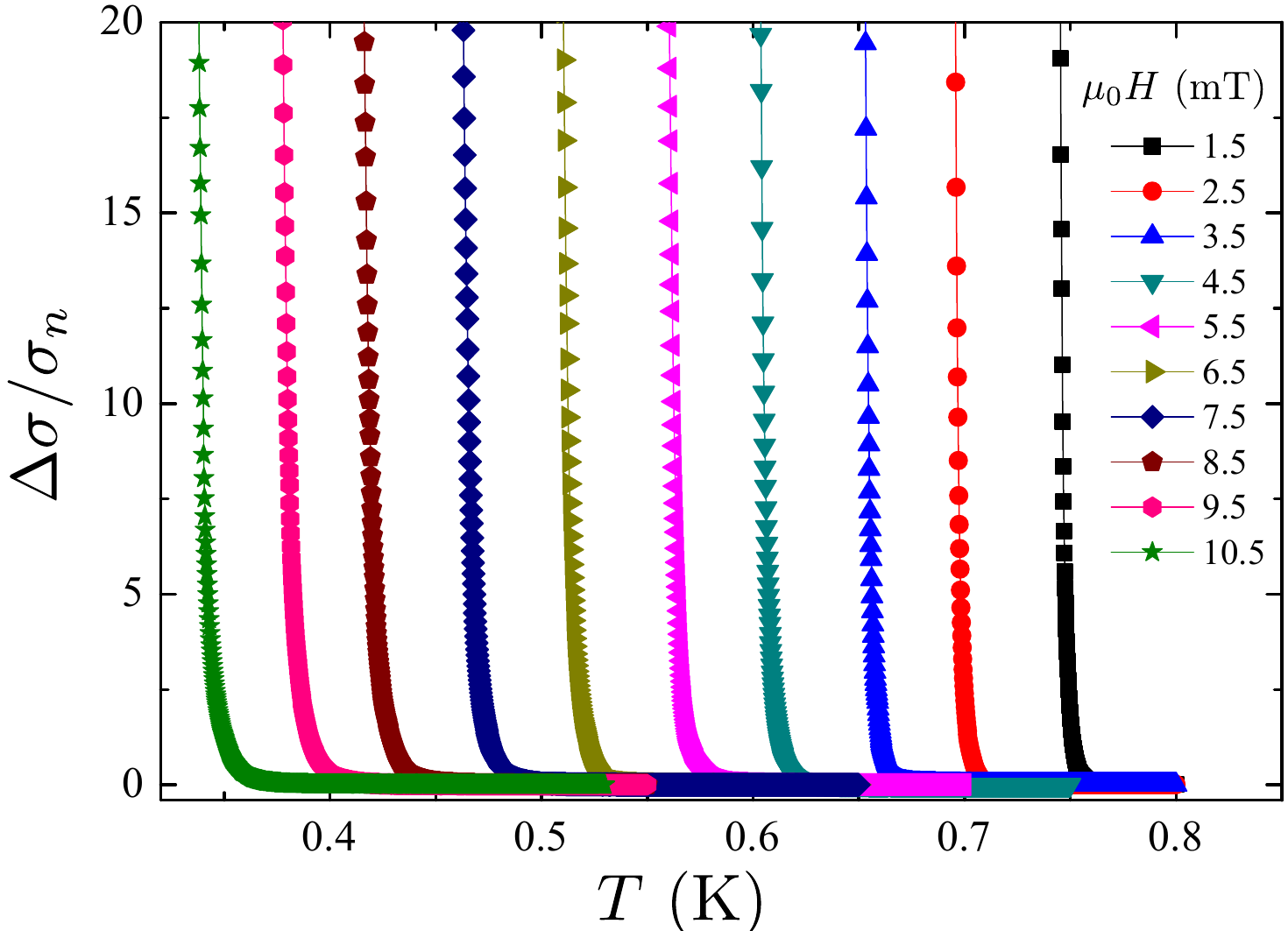}
\setlength{\belowcaptionskip}{-12pt}\setlength{\abovecaptionskip}{-1pt}
\caption{
The normalized excess conductivity plotted as a function of the temperature in a magnetic field $\mu_0 H$ ranging from $1.5$ to \SI{10.5}{mT}.
}
\label{EXC}
\end{figure}

\Cref{EXC} shows the normalized excess conductivity $\Delta\sigma_{H}(T)/\sigma_n$ curves for a WP single crystal in applied magnetic field up to \SI{10.5}{mT} and with direction parallel to the $a$-axis. 
In particular, the excess conductivity due to fluctuation effects near the superconducting transition is defined as $\Delta\sigma_{H}(T)=\sigma_{H}(T)-\sigma_n(T)$, with $\sigma_H(T)$ the sample conductivity and $\sigma_n(T)$ the normal state conductivity. 
The data has been obtained by the excess conductance calculated as $\Delta\Sigma_H(T)=\Sigma_H(T)-\Sigma_{n}(T)$, with $\Sigma_H(T)$ the measured conductance and $\Sigma_{n}(T)$ the normal state conductance. 
In the temperature range investigated the normal conductance is temperature-independent $\Sigma_{n} =\SI{41}{\Omega^{-1}}$. 

\begin{figure}
\centering
\includegraphics[width=\columnwidth]{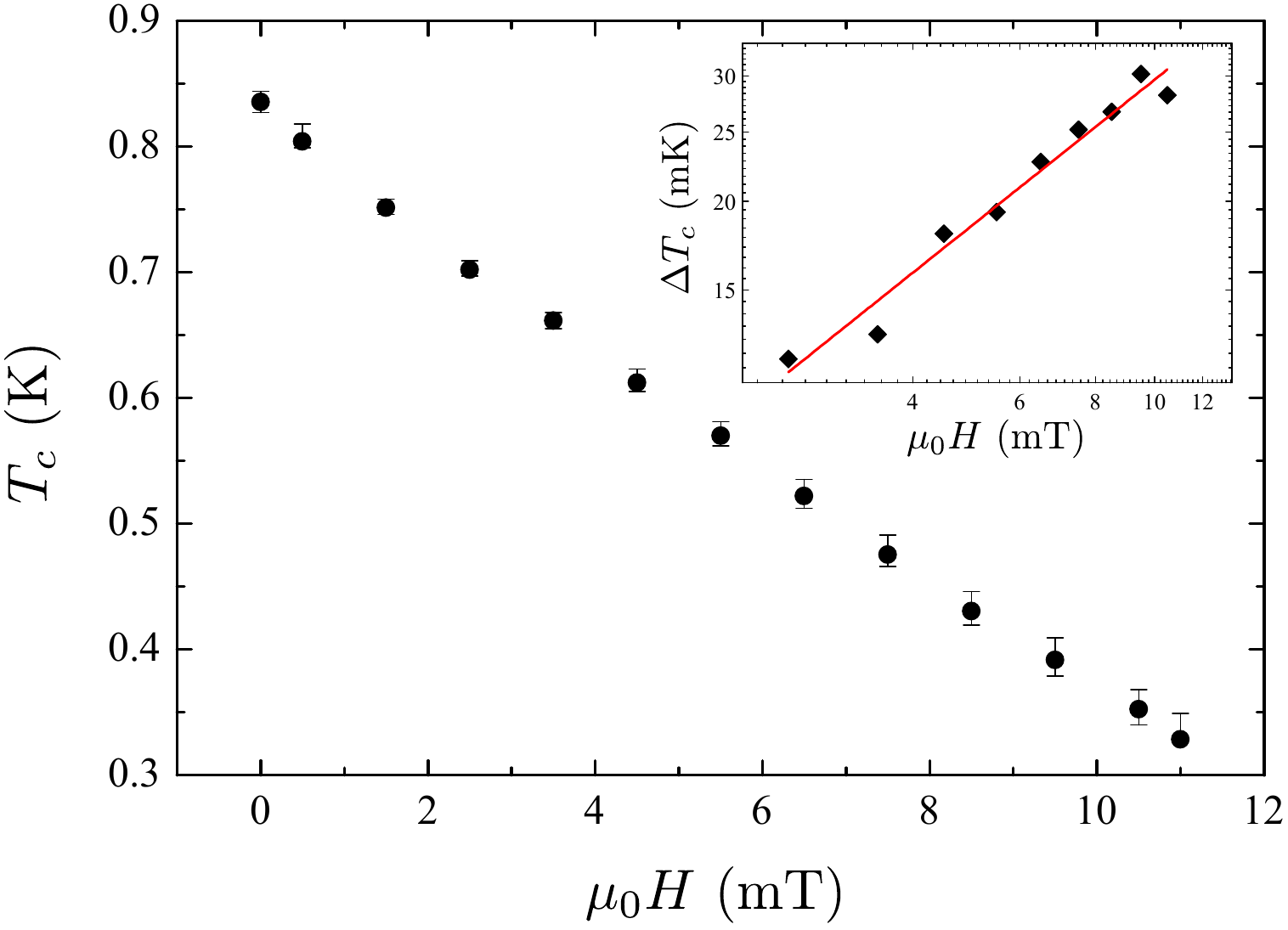}
\setlength{\belowcaptionskip}{-12pt}\setlength{\abovecaptionskip}{-1pt}
\caption{
The critical temperature as a function of the applied magnetic field $\mu_0 H$. 
Error bars correspond to the transition width $\Delta T_c$ evaluated from the 10--90\% criterion. 
The inset refers to the transition width $\Delta T_c$ versus $\mu_0 H$. The solid red line is the linear best fit to the experimental data.
}
\label{TCH}
\end{figure}

\cref{TCH} shows the critical temperature $T_c$ as a function of the applied magnetic field. The errors bars correspond to the transition width determined from the 10--90\% resistance drop criterion. The inset shows the transition width $\Delta T_c$ obtained by the 10--90\% criterion as a function of the applied magnetic field on a log-log scale. 
A power-law behavior $H^{-\alpha}$ is inferred, with an exponent $\alpha=0.68$ very close to the value $2/3$ predicted for field induced fluctuation effects in a 3D superconductor. 
Therefore, this result suggests a 3D behavior of the conductivity fluctuations.

\begin{figure}
\centering
\includegraphics[width=\columnwidth]{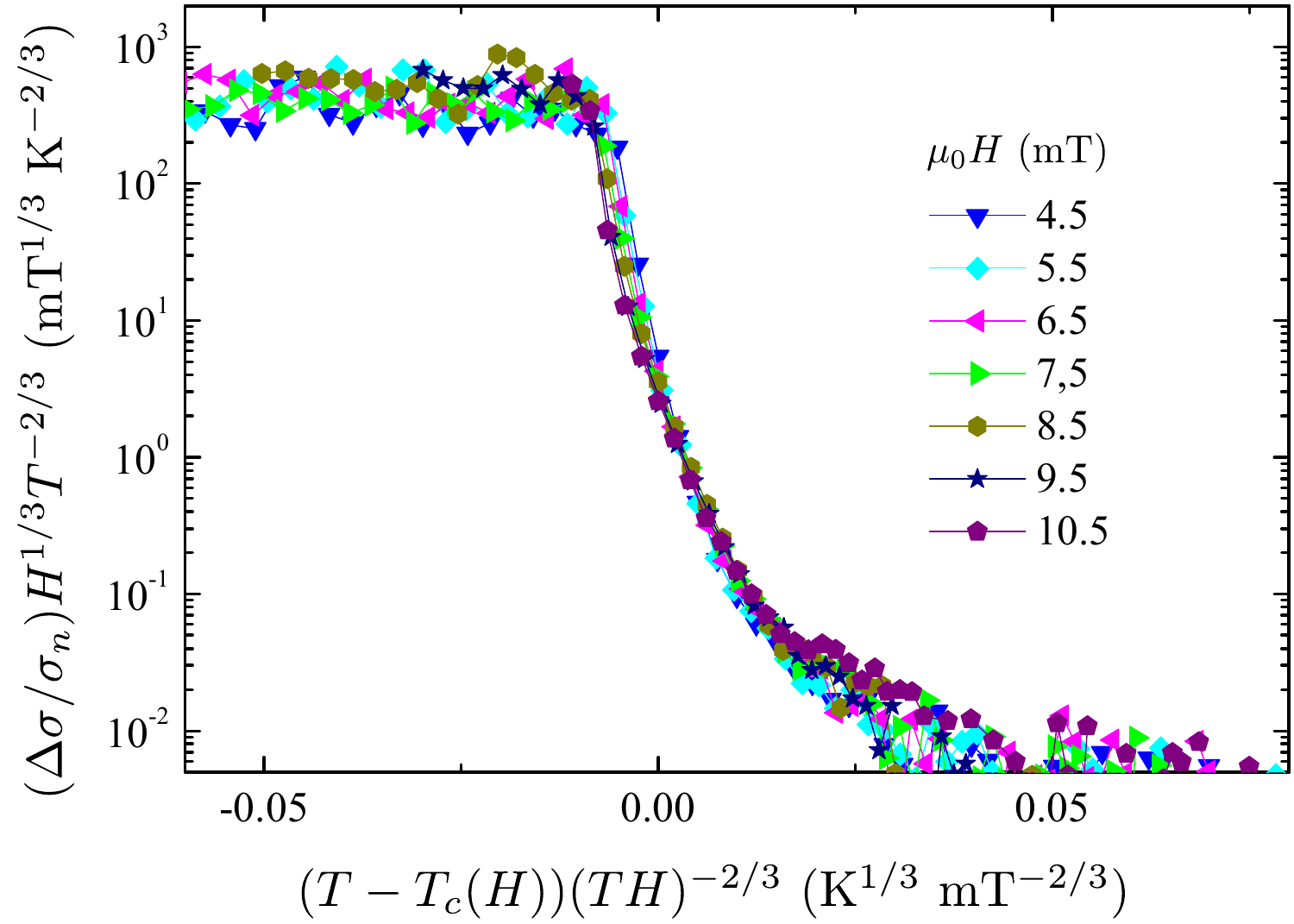}
\setlength{\belowcaptionskip}{-12pt}\setlength{\abovecaptionskip}{-1pt}
\caption{
Scaling plots of $(\Delta\sigma/\sigma_n)(H^{1/3} T^{-2/3})$ as a function of $(T-T_c(H))/(TH)^{2/3}$ for the transition curves in \cref{EXC} at magnetic fields $\geq\SI{4}{mT}$, 
for the 3D Ullah-Dorsey model of the paraconductivity described by \cref{UD3D}.
}
\label{SCALING}
\end{figure}

In \cref{SCALING} we plot the scaled excess conductivity $\Delta\sigma_{3D}(H)$ for the case of 3D scaling. 
Notice that the 3D scaling behavior has been calculated from the normalized excess conductivity $\Delta \sigma_H(T)/\sigma_n$ curves shown in \cref{EXC}, using $T_c(H)$ values shown in \cref{TCH}. 
For fields $\geq\SI{4}{mT}$, all data points corresponding to the different superconducting transitions collapse onto a single curve, thus exhibiting a reasonable scaling behavior of the fluctuations around $T_c(H)$.
Hence, this result indicates that, at sufficiently high fields, the fluctuation conductivity is well described within the 3D LLL approximation and that the field $\mu_0 H_{LLL}$ is of the order of \SI4{mT}.

As noted before, the relevant parameter that quantifies the fluctuation strength in a superconductor is the Ginzburg number given by \cref{G0}. In order to estimate the 3D Ginzburg number for the WP, the thermodynamic critical field $\mu_0 H_c(0)$ may be inferred by the jump of specific heat at $T_c$ given by $\mu_0 \Delta C / T_c= \left ( \mu_0 H_c(0)/T_c \right )^2$, whereas the in-plane coherence length $\xi_\para$ by the slope near $T_c$ of the out-of-plane upper critical field, $dH^{\perp}_{c2}/dT|_{T=T_c}$ given by 
\begin{equation}
\xi_\para= \sqrt{\Phi_0\Big/ \left (2\pi T_c\mu_0 \left |\frac{dH^{\perp}_{c2}}{dT}\Big| _{T=T_c} \right | \right )}.
\end{equation}
In terms of the measured quantities $\mu_0 \Delta C / T_c$, $dH^{\perp}_{c2}/dT|_{T=T_c}$, $\Gamma$, and $T_c$, the Ginzburg number can be expressed as
\begin{equation}
G^\mathrm{3D}_i(0)=2 \left ( \frac{k_B \Gamma}{\Delta C} \right )^2 \left ( \frac{2 \pi T_c \mu_0}{\Phi_0} \frac{dH^{\perp}_{c2}}{dT}\Big| _{ T=T_c} \right )^3\, .
\end{equation}
The specific heat jump $\mu_0 \Delta C / T_c=\SI{120}{J/m^3K^2}$ of WP single crystal has been already measured and reported elsewhere~\cite{Liu19}, while the slope $dH^{\perp}_{c2}/dT|_{T=T_c}=\SI{-22}{mT/K}$ has been obtained by the $H^{\perp}_{c2}(T)$ extracted from the data in the inset of \cref{SCALING} and $\Gamma \approx 1.4 $. Thus, we found that $G^\mathrm{3D}_i(0)\approx 10^{-8}$, 
which is very small compared, for instance, with the value $\approx 10^{-2}$ observed in the iron-selenide superconductor~\cite{Hardy20}, whereas it is comparable to the value observed in the low-temperature superconductor niobium~\cite{Blatter94,Koshelev19,Liarte17}.
As stated before, the Ginzburg number $G_i$ measures the strength of thermal fluctuations at the superconducting transition. 
In particular, it determines the width of the temperature interval around the critical temperature $T_c$ in which fluctuations effects are observable and affects several features of the H-T phase-diagram, e.g., the vortex melting line~\cite{Blatter94}. 
Large anisotropy parameters and high critical temperatures correspond to larger Ginzburg numbers $G_i \propto (\Gamma T_c)^2$. 
In conventional superconductors, $G_i \approx 10^{-8}$ as for niobium~\cite{Blatter94,Koshelev19,Liarte17,Forgan02}. 
In high-temperature cuprates and iron-based superconductors, $G_i$ is up to $10^{-2}$. 
The obtained value $10^{-8}$ for WP suggests that thermal fluctuations are comparable to those of conventional superconductors. 
This also suggests a H-T phase-diagram similar to those observed for conventional superconductors, which is mainly affected by disorder and with a vortex phase less affected by thermal fluctuations~\cite{Mikitik01,Villegas05}. 

Another relevant parameter describing the superconducting state is the Ginzburg-Landau parameter $\kappa$ given by
\begin{equation}
\kappa= 
\frac{\mu_0}{\sqrt{2\mu_0 \Delta C/T_c}} 
\frac{dH^{\perp}_{c2}}{dT}\Big|_{T=T_c} ,
\end{equation}
that for our sample is $\kappa\approx 1.3$, which is again of the same order of magnitude of the value measured for niobium~\cite{Koshelev19,Liarte17}.

\section{Conclusion}

In conclusion, we have synthesized superconducting single crystals of WP, and investigated electrical transport properties. The synthesis of WP single crystals was accomplished through the chemical vapor transport method which has been proved to be successful to grow transition metal pnictides. 
Our analysis shows that the angular dependence of the upper critical field exhibits a smooth behavior. Looking at the normal state, we extract a rather large anisotropy, while in the superconducting state, the upper critical field shows an anisotropy $\Gamma=1.4$, largely lower than that found, for instance, in iron sulfides~\cite{Borg16} and organic superconductors~\cite{Yamamoto18}. We note that this value for $\Gamma$ corresponds to an estimated effective mass anisotropy equal to ${m_{\perp}^*}/{m_{\para} ^*}\approx2$. Moreover, the magnetoresistance measurements performed at different applied magnetic field angles reveal a 3D behavior differently from the 2D character found in iron selenide~\cite{Sun16}. On the other hand, the fit of the superconducting fluctuations of the conductivity, by means of Ullah-Dorsey theory, suggests again a 3D scaling law rather than a 2D behavior. 
Therefore, these experimental data, supplemented by the theoretical theories used to fit their trend, indicate that the WP can be considered an anisotropic 3D superconductor.

It is worth stressing that these results are corroborated by ab-initio electronic structure calculations that show anisotropic hopping parameters, whose values clearly indicate a 3D behavior. 
Interestingly, most of the density of states at the Fermi energy is contributed by W-5$d$ electrons, also suggesting that the superconductivity is originated from the condensation of electrons coming from the transition metal ion. 

Nevertheless, further theoretical and experimental studies are needed to determine the pairing symmetry and the corresponding superconducting mechanism and the role played by W-5$d$ electrons in stabilizing the superconducting phase.

\section*{Acknowledgments}

P.~M.~is supported by the Japan Science and Technology Agency (JST) of the Ministry of Education, Culture, Sports, Science and Technology (MEXT), JST CREST Grant No.~JPMJCR19T and by the Japan Society for the Promotion of Science (JSPS) Grant-in-Aid for Early-Career Scientists (Grant No.~20K14375). 
C.~A. and G.~C. are supported by the Foundation for Polish Science through the International Research Agendas program co-financed by the European Union within the Smart Growth Operational Programme.
C.~A. and G.~C. acknowledge the access to the computing facilities of the Interdisciplinary Center of Modeling at the University of Warsaw, Grant No.~GB84-7.
We acknowledge the CINECA award under the ISCRA initiative IsC81 "DISTANCE" Grant, for the availability of high-performance computing resources and support.


\end{document}